# Fluctuations, Response, and Resonances in a Simple Atmospheric Model

Andrey Gritsun[1,2] and Valerio Lucarini[3,4,5]

1. Institute of Numerical Mathematics, Russian Academy of Sciences, Moscow, Russian Federation
2. Institute of Applied Physics, Russian Academy of Sciences, Nizhny Novgorod, Russian Federation
3. Department of Mathematics and Statistics, University of Reading
4. Centre for the Mathematics of the Planet Earth, University of Reading
5. CEN, University of Hamburg, Hamburg, Germany

February 22nd 2017

**Abstract**

We study the response of a simple quasi-geostrophic barotropic model of the atmosphere to various classes of perturbations affecting its forcing and its dissipation using the formalism of the Ruelle response theory. We investigate the geometry of such perturbations by constructing the covariant Lyapunov vectors of the unperturbed system and discover in one specific case – orographic forcing – a substantial projection of the forcing onto the stable directions of the flow. This results into a resonant response shaped as a Rossby-like wave that has no resemblance to the unforced variability in the same range of spatial and temporal scales. Such a climatic surprise corresponds to a violation of the fluctuation-dissipation theorem, in agreement with the basic tenets of nonequilibrium statistical mechanics. The resonance can be attributed to a specific group of rarely visited unstable periodic orbits of the unperturbed system. Our results reinforce the idea of using basic methods of nonequilibrium statistical mechanics and high-dimensional chaotic dynamical systems to approach the problem of understanding climate dynamics.

1. **Introduction**

Nowadays statistical mechanics and thermodynamics provide a comprehensive picture of the properties of equilibrium and near-equilibrium systems[1,2]. On the contrary, our understanding of nonequilibrium systems is comparatively poor and limited, despite the wealth of phenomena that are possible only far from equilibrium conditions. Advancing our knowledge on nonequilibrium systems is one of the great frontiers of contemporary science[3]. Conceptually, one can represent a large class of nonequilibrium systems as being in contact with at least two reservoirs having different temperature (and/or chemical potential), with a nonequilibrium steady state emerging after transients have died out[4]. The presence of nonequilibrium conditions is essential for sustaining, e.g., life or convective motions, through processes allowing for a systematic conversion of energy from one form to another one, like a heat engine. Additionally, irreversible processes lead to the generation of entropy, which is transferred into the reservoirs[5].

The climate is a prototypical example of a nonequilibrium, forced, and dissipative system, where the unequal absorption of solar radiation triggers a wealth of processes and feedbacks leading to the presence of a variability in the climatic fields covering a large range of scales in space and in time[6]. As a result, our ability to observe the climate state is intrinsically limited. It is extremely challenging to provide a convincing and comprehensive framework underpinning climate variability and able to predict its response to perturbations. This would be great importance for reconstructing the climate of the past and predicting its response to forcings such as changes in the atmospheric composition, land surface properties, incoming solar radiation, and position of the continents[7].

An extremely useful point of view on climate is given by thermodynamics: the organized motions of the geophysical fluids result from the transformation of available potential into kinetic energy, which is eventually dissipated by friction[7,8]. At this regard, one can construct a precise analogy between the functioning of the climate system and an imperfect, irreversible engine, characterized by the presence of positive correlations between heating patterns and temperature anomalies. In turn, the atmospheric winds and the oceanic currents contribute to decreasing the temperature and chemical potential (mostly due to inhomogeneity of water vapour – in the atmosphere - and salinity – in the ocean) gradients that fuel such motions: this provides stabilizing negative feedbacks that make it possible, together with the global Boltzmann radiative feedback, for the climate to reach a steady state[9].

In the case of (near) equilibrium systems, one can predict the impact of applying a (weak) forcing using the statistics of the unperturbed system through the fluctuation-dissipation theorem (FDT), which establishes a dictionary (in the form of linear operators) for translating natural into forced fluctuations, and viceversa[10,11]; extensions have been recently proposed for the nonlinear case[12].

Correspondingly, in order to bypass many of the bottlenecks mentioned above, it is very tempting to try to relate climate variability to its response to external forcings. In the case of Axiom A systems[13], it is indeed possible to construct a rigorous response theory, able to describe the change in the measure resulting from perturbing the system in terms of the properties of the unperturbed measure[14,15,16]. Nonetheless, in nonequilibrium conditions there is no obvious relationship between free and forced fluctuations of the system[17], as already suggested by Lorenz[18].

Physically, one can interpret such a lack of equivalence as resulting from the fact the natural fluctuations explore only the unstable manifold in the tangent space, while the forced fluctuations can explore both the stable and unstable directions. Mathematically, this has to do with the fact that the invariant measure is smooth only along the unstable directions of the flow. We remark that mathematical results strictly valid for Axiom A systems are considered extremely relevant for general high dimensional chaotic physical systems because of the chaotic hypothesis[19].

In order to bypass the problem of the lack of smoothness of the invariant measure and be able to take advantage of the FDT, various points of view have been proposed. In particular, some authors consider adding some stochastic forcing to the deterministic dynamics, so to consider the impact of unresolved scales[20]; see also discussion in [21,22]. This gives a rationale for using the FDT in the context of the climate models: while on one side there have been successful examples of applications of the FDT, predictive power depends substantially on the chosen observable[23,24,25].

The ab-initio implementation of the Ruelle response operators is hindered by the presence of differing behavior between the terms contributing along the stable and unstable directions of the flow[26]. Algorithms based on adjoint methods seem to partially ease these issues[27]. A possible idea for improving the convergence of the algorithm relies on projecting the perturbation flow on the unstable, neutral and stable covariant Lyapunov vectors[28](CLVs), which have been recently constructed for rather nontrivial geophysical fluid dynamical systems[29,30]. Using a different approach, based on exploiting some formal properties and taking advantage of a set of test simulations, the Ruelle response theory has been shown to provide a considerable degree of predictive power in systems ranging from simple low dimensional models[31,32,33] to very high dimensional climate models with hundreds of thousands of degrees of freedom[9,34,35].

An alternative point of view on the problem can also be taken. In the case of Axiom A systems, an infinite set of unstable periodic orbits (UPOs) populate densely the attractor, and it is possible to evaluate the expectation value of a measurable observable as an average over the various UPOs, each taken with a suitable weight. The weights are smaller for more unstable orbits, so that the least unstable orbits give the largest contributions, and provide a natural way to reconstruct hierarchically the properties of the attractor [36,37,38]. While intuition seems to suggest that the numerics of constructing closed unstable orbits in chaotic systems (especially high-dimensional ones) is hopeless, this is indeed not the case: UPOs are used as practical tools for studying, interpreting, and reconstructing chaotic dynamical systems of various degrees of complexity, see[39].

UPOs have been constructed also in the case of simple yet relevant models of the atmosphere[40,41,42] and provide the framework of potentially addressing a classic problem of climate dynamics and dynamical meteorology, i.e. the identification of so-called modes or regimes, whose investigation has started through the classic concept of Grosswetterlagen[43]. More comprehensive approaches and definitions have been later given in the 1980s[44,45,46]; see an illuminating discussion in [47], while in [48] one can find many details on how to study and detect such regimes through time series analysis. The idea is that the least unstable UPOs might provide the key elements for reconstructing and understanding weather and climate regimes, and might be directly related to the principal modes of the system found using linear [48] and nonlinear [49] empirical methods, while transitions between the weather regimes could be interpreted as transitions between neighborhoods of UPOs[50]. This has crucial relevance in terms of performing extended prediction of the state of the climate system. We will address these specific research questions elsewhere.

The overall goal of this paper is to show the potential of some basic ideas of nonequilibrium statistical mechanics and high-dimensional chaotic dynamical systems to approach the problem of understanding climate dynamics. Using the mathematical framework of Axiom A dynamical system and taking advantage of the Ruelle response theory, we study the change in the statistical and dynamical properties of a simple quasi-geostrophic barotropic model[51,52] of the northern hemisphere atmosphere in a strongly chaotic regime, first described in [40], resulting from two different kinds of perturbations – namely, we shall change the orography, and the boundary layer dissipation of the system.

We anticipate some of the main results for the benefit of the reader. We find a relevant example of a case where a fundamental property on nonequilibrium system, i.e. the nonequivalence between forced and free motions, is apparent. We discover a resonance occurring in a frequency range where the unperturbed system does not have virtually any signal in terms of power spectrum. The resonance can be associated via UPOs analysis to an orographically forced Rossby-like wave, which corresponds to a set of extremely unstable UPOs. The analysis of CLVs shows that, indeed, a substantial part of the signal is associated to perturbed motions forced along the stable direction of the flow, so that they cannot be captured within the natural variability of the system. There is, in fact, no sign of such a Rossby-like wave within the variability of the unperturbed flow. Such a *climatic surprise* is a clear example in a high-dimensional nonequilibrium system of conditions under which the FDT can fail. Indeed, we show that the performance of the FDT in reconstructing the response of the system is much worse in the case of perturbations to the orographic forcing than in the case of perturbations to the boundary layer friction, where instead the role of the stable directions in the tangent space is less prominent.

The paper is structured as follows. In Sect. 2 we provide a recapitulation of the main mathematical concepts relevant for the interpretation of the results presented in this paper, in order to make it as self-contained as possible. In Sect. 3 we describe the atmospheric models used in this study. In Sect. 4 we present the numerical experiments and describe the results of our investigation. In Sect. 5 we give a summary of our main findings and draw our conclusions. Finally, in Appendix

A we present a short summary of the statistical tools used for analyzing the outputs of the numerical model.

## 2. The Mathematical Setting

*2.1 Axiom A Dynamical Systems*

Following the chaotic hypothesis, Axiom A dynamical systems provide a good mathematical framework for general chaotic physical systems[19]. We then consider a continuous-time autonomous Axiom A dynamical system $\dot{x} = f(x)$ on a compact manifold $\mathcal{M} \subset \mathbb{R}^n$, where $x(t) = S^t x_0$, with $x_0 = x(0)$ initial condition and $S^t$ evolution operator defined for $t \in \mathbb{R}_{\geq 0}$. Let us define $\Omega \subset \mathcal{M}$ as the compact attracting invariant set of the dynamical system, so that we define $\rho$ as the associated Sinai-Ruelle-Bowen (SRB) measure[13] with support $\Omega = supp(\rho)$. The SRB measure is rather satisfactory in terms of physical intuition as it can be characterized as being the zero-noise limit of the invariant measures obtained when adding a small stochastic forcing of vanishing amplitude on top of the deterministic dynamics given by $\dot{x} = f(x)$. We have that the dynamical system is uniformly hyperbolic for $x \in \Omega$ and is ergodic, so that, for some generic observable $\Phi$

$$\rho(\Phi) = \langle \Phi \rangle_0 = \int \rho\,(dx)\Phi(x) = \lim_{t \to \infty} \frac{1}{t} \int_0^t d\tau\, \Phi(S^\tau x) \quad (1)$$

for almost all (in the Lebesgue sense) initial conditions $x$ belonging to the basin of attraction of $\Omega$.

*2.2 Lyapunov Exponents and Covariant Lyapunov Vectors*

For Axiom A systems (and, in fact, also for more general ones) it is possible to construct $n$ Lyapunov exponents (LEs) $\lambda_1 \geq \lambda_2, \dots, \geq \lambda_n$, which describe the asymptotic behavior of infinitesimal perturbations (exponential growth or decay) from a typical background trajectory. See a comprehensive treatment in [53]. As well known, the presence of at least one positive Lyapunov exponent indicates that the evolution of the system has sensitive dependence with respect to the initial conditions and is taken as one of the possible definitions of chaos. We have that $\sum_{i=1}^n \lambda_i = \int \rho\,(dx) \nabla_x \cdot f(x)$, i.e. the sum of the Lyapunov exponents is equal to the expectation value of the divergence of the flow. Instead, the Kolmogorov-Sinai entropy $h_{ks} = \sum_{\lambda_i > 0} \lambda_i$ measures the rate of production of information of the dynamical system and provides a characterization of its limited predictability. Axiom A systems describing nonequilibrum, forced and dissipative systems feature a negative sum of their Lyapunov exponents, thus implying that the phase space contracts in time. Therefore, the *n*-dimensional Lebesgue measure of $\Omega$ vanishes and one can introduce generalized notions of (fractal) dimension in order to provide quantitative characterizations of $\Omega$. While the theory of Renyi dimensions gives an overarching method to study the geometrical properties of strange geometrical objects, the Kaplan-Yorke conjecture says that it is possible to provide a meaningful definition of the fractal dimension of $\Omega$ using the spectrum of the Lyapunov exponents as follows[53]:

$$d_{KY} = m + \frac{\sum_{i=1}^{m} \lambda_i}{|\lambda_{m+1}|} \leq n \quad (2)$$

where $m$ is the largest number such that $\sum_{i=1}^{m} \lambda_i \geq 0$. If the system is not dissipative, we have that $d_{KY} = n$. Note that for Axiom A systems all of the generalized Renyi dimensions of $\Omega$ have the same value [54].

It is possible to supplement the notion of LEs with a powerful geometric construction. The Covariant Lyapunov Vectors (CLVs) provide a covariant basis $\{c_1(t), c_2(t), \ldots, c_n(t)\}$ describing the solutions to the following system of linear ordinary differential equations:

$$\dot{y} = M(S^t x)y \quad (3)$$

where $M(y) = \nabla_y f(y)$ is the tangent linear. The main property of the basis of CLVs is that setting $c_j(t_1)$ as initial condition for $y$ at time $t_1$ in Eq. (3), at time $t_2 > t_1$ the solution is parallel to $c_j(t_2)$, and, in the limit for $t_2 \to \infty$, the average rate of growth or decay of its amplitude is the $j^{th}$ LE $\lambda_j$. Therefore, the CLVs provide explicit information about the directions of asymptotic growth and decay in the tangent linear space. The CLVs corresponding to positive (negative) LEs span the unstable (stable) tangent space. The CLVs corresponding to the vanishing LE is oriented along the direction of the flow and spans the neutral direction of the tangent space. Efficient algorithms for identifying the CLVs were first determined independently in [28] and [55] by a suitable combination of calculations of the linear spaces constructed when computing the Lyapunov exponents. See a comprehensive review in [56].

*2.3 Response Theory*

Axiom A systems provide a suitable framework for constructing a rigorous response theory, see a review in [16] for the point of view proposed here and [57] for a different approach based on the transfer operator formalism. We perturb the evolution equation $\dot{x} = f(x)$ having invariant measure $\rho$ by adding a weak perturbation, so that the new dynamics is described by

$$\dot{x} = f(x) + \varepsilon X(x)g(t), \quad (4)$$

where $X(x)$ is a smooth vector field defining the pattern of the forcing in the phase space, $g(t)$ is the time modulation of the forcing, and $\varepsilon$ is the small parameter controlling the strength of the perturbation. Ruelle showed that the expectation value of a sufficiently smooth observable $\Phi$ in the forced system can be computed as a perturbative expansion with respect to powers of $\varepsilon$ as $\langle \Phi \rangle_{g,X,\varepsilon}(t) = \langle \Phi \rangle_0 + \delta \langle \Phi \rangle_{g,X,\varepsilon}(t)$, where $\delta \langle \Phi \rangle_{g,X,\varepsilon}(t) = \sum_{n=1}^{+\infty} \varepsilon^n \langle \Phi \rangle_{g,X}^{(n)}(t)$ is the correction term resulting from the perturbation, expressed as a series. In particular, the first term in the series gives

the linear response:

$$\langle\Phi\rangle_{g,X}^{(1)}(t) = \int_{-\infty}^{+\infty} d\sigma_1 G_{\Phi,X}^{(1)}(\sigma_1)\, g(t-\sigma_1) \quad (5)$$

where $G_{\Phi,X}^{(1)}(t)$ is the first order Green's function:

$$G_{\Phi,X}^{(1)}(\sigma_1) = \Theta(\sigma_1) \int \rho\,(dx) X(x) \cdot \nabla_x \Phi(S^{\sigma_1} x) \quad (6)$$

where, notably, both the averaging described by $\rho$ and the evolution operator $S^\tau$ refer to the unperturbed dynamics, and $\Theta$ is the Heaviside distribution. Note that $G_{\Phi,X}^{(1)}(t)$ is a causal function, so that $G_{\Phi,X}^{(1)}(t) = 0$ if $t < 0$. Under standard condition of integrability, by taking the Fourier transform of $\langle\Phi\rangle_{g,X}^{(1)}(t)$ we derive

$$\widetilde{\langle\Phi\rangle}_{g,X}^{(1)}(\nu) = \chi_{\Phi,X}^{(1)}(\nu)\tilde{g}(\nu) \quad (7)$$

where the linear susceptibility $\chi_{\Phi,X}^{(1)}(\nu) = \int_{-\infty}^{\infty} dt\, G_{\Phi,X}^{(1)}(t) \exp[i2\pi\nu t]$ is the Fourier transform of $G_{\Phi,X}^{(1)}(t)$, and $\tilde{g}(\nu)$ is the Fourier transform of $g(t)$. The susceptibility describes the response of the system to forcings at different frequencies. The real and imaginary parts of the function represent the in- and out-of-phase response of the system respectively to a sinusoidal forcing at frequency $\nu$. Maxima in the absolute value of the susceptibility correspond to resonances of the system. Note that the susceptibility $\chi_{\Phi,X}^{(1)}(\nu)$ obeys the following identity [58]:

$$\chi_{\Phi,X}^{(1)}(\nu) = \frac{i}{\pi} P \int_{-\infty}^{\infty} d\nu' \frac{\chi_{\Phi,X}^{(1)}(\nu')}{\nu' - \nu} \quad (8)$$

where $P$ indicates integration in principal part, and the susceptibility is related to its complex conjugate through $\chi_{\Phi,X}^{(1)}(\nu) = [\chi_{\Phi,X}^{(1)}(-\nu)]^*$. Equation (8) can be recast in terms of conventional Kramers-Kronig relations (KK), linking the real and imaginary parts of $\chi_{\Phi,X}^{(1)}(\nu)$, see [17, 32, 59].

As discussed in [16], since in the case of dissipative systems the SRB measure $\rho$ is absolutely continuous with respect to Lebesgue along the unstable directions, but not so along the stable directions of the tangent space, one cannot rewrite the Green's function (6) in the form of a fluctuation-dissipation theorem (FDT). This means that forced and free fluctuations are not equivalent in a nonequilibrium system, as opposed to the equilibrium case; see discussion in [9, 32]. In practice, this implies that, taking an observable $\Phi$, the squared modulus of its susceptibility $\chi_{\Phi,X}^{(1)}(\nu)$ might be radically different from the power spectrum of $\Phi$ in the unperturbed state.

If one assumes that the unperturbed invariant measure is Gaussian, so that $\rho(dx) = \rho_0 \exp(-(x-\langle x\rangle_0)^T \bar{C}^{-1}(x-\langle x\rangle_0)/2)\,dx$, where $\bar{C}$ is the covariance matrix, $\rho_0$ is a normalizing factor, and $dx$ is the Lebesgue measure in $\mathbb{R}^n$, the Green's function takes the following form:

$$G^{(1)}_{\Phi,X}(\sigma_1) = \Theta(\sigma_1) \int \rho(dx) \, ((x - \bar{x})^T \bar{C}^{-1} X(x) - \nabla_x \cdot X(x)) \Phi(S^{\sigma_1} x). \quad (9)$$

The previous expression further simplifies if the forcing is divergence free (and in particular state-independent), so that $\nabla_x \cdot X(x) = 0$; in this case the classical Kubo-like FDT is recovered. If one approximates the unperturbed invariant measure of a system with its Gaussian fit, Eq. (9) can be used for estimating the Green's function using the so-called quasi-Gaussian FDT (qG-FDT) approximation.

## *2.4 Unstable Periodic Orbits*

One of the properties of an Axiom A system is that its attractor is densely populated by unstable periodic orbits (UPOs)[37,38,39], which provide the so-called skeletal dynamics of the system. Therefore, one can formally construct the invariant measure $\rho$ of the system by considering the following expression for the expectation value of any measurable observable $\Phi$:

$$\rho(\Phi) = \lim_{t \to \infty} \frac{\sum_{\Pi^p, p \leq t} w^{\Pi^p} \overline{\Phi^{\Pi^p}}}{\sum_{\Pi^p, p \leq t} w^{\Pi^p}} \quad (10)$$

where $\Pi^p$ is the UPOs of prime period $p$, $\overline{\Phi^{\Pi^p}}$ is the average of the observable $\Phi$ taken on the orbit $\Pi^p$, and $w^{\Pi^p}$ is the weight of the UPO $\Pi^p$, which depends on, among other things, how unstable $\Pi^p$ is. A detailed discussion of Eq. (10) can be found in [39] and [60]. The previous expression clarifies that a resonance for $\nu = \nu_0$ in the susceptibility $\chi^{(1)}_{\Phi,X}(\nu)$ may be related to the existence of one or more UPOs having period $p \approx 1/\nu_0$. The analysis of the corresponding UPOs might shed light on the resonant mechanisms, compare, e.g., results in [31] and [61].

## 3. The Atmospheric Model

The goal of this paper is to advance our understanding of the statistical mechanics of climate by studying its variability and response to forcings. Since we are testing conceptually non-trivial methods, it is necessary to choose a simplified model of the climate. We consider a simple quasi-geostrophic barotropic model of atmosphere of the northern hemisphere, featuring forcing and dissipation. The geophysical fluid dynamical basics behind the quasi-geostrophic approximation, the basic features of barotropic flows, and the historical relevance of such a class of models can be found in [51,52] and will not be discussed here. Despite its simplicity, the model can be tuned in such a way to have a high dimensional attractor and statistical characteristics resembling the basic features of the observed large scale atmospheric circulation.

Defining $\Psi$ as the streamfunction, such that the horizontal wind velocity is given by its orthogonal (i.e. rotated clockwise by $\pi/2$) gradient as $(u, v) = \nabla^\perp \Psi$ (no vertical wind velocity exists in this model), the quasi-geostrophic barotropic model model solves the evolution equation for the relative vorticity $\nabla^2 \Psi$ ($\nabla^2$ is the Laplacian operator) in the rotating spherical surface

$$\frac{\partial \nabla^2 \Psi}{\partial T} + J(\Psi, \nabla^2 \Psi + L + K_H H) = -A\nabla^2 \Psi + M\nabla^4 \Psi + F_{ext} \qquad (11)$$

Our model takes into account the modifications to the vorticity equations resulting from the orographic forcing and the beta effect. These are represented by the terms $K_H H$ and $L = 2\Omega_E \sin \vartheta$ inside the Jacobian operator $J(.,.)$, where $H$ is the orography, $K_H = \sqrt{2}\Omega_E/H_A$, $\Omega_E$ is the Earth angular velocity, $H_A$ is the standard height of the atmosphere and $\vartheta$ is the latitude. As dissipative terms, we consider the effect of the boundary layer giving an Ekman friction term of the form $-A\nabla^2\Psi$, plus the effect of turbulent diffusion included via the term $M\nabla^4\Psi$; the dissipation coefficients A and M will be specified later. In order to inject energy into the system, we also include a constant forcing $F_{ext}$ that provides a severely simplified parametrization of the net effect of baroclinic modes and of unresolved small scale barotropic modes. The description of how the forcing is constructed comes few paragraphs below.

We then transform Eq. (11) into dimensionless form using the inverse of the Earth angular velocity $\Omega_E$ and the Earth radius $R_E$ as time and length scales, so that $t$ and $\Psi$ units are $1/\Omega_E$ and $R_E^2/\Omega_E$, respectively, and, after applying an inverse Laplacian operator, we obtain:

$$\frac{\partial \psi}{\partial t} + (\nabla^2)^{-1} J(\psi, \nabla^2 \psi + l + k_H h) = -\alpha \psi + \mu \nabla^2 \psi + f_{ext}. \qquad (12)$$

Here $\psi, h, l, f_{ext}$ are the dimensionless streamfunction, orography, Coriolis term ($l = 2\sin\vartheta$) and external forcing. Next, we project Eq. (12) via standard Galerkin methods using the basis of spherical harmonics $\{Y_i, i = 1..N_d\}$ antisymmetrical with respect to equator (our domain is limited to the northern hemisphere) and obtain the finite dimensional approximation of (12). This is then implemented as a computer code. We adopt a T21 truncation in Galerkin method with the total number of degrees of freedom of the model is $N_d = 231$. Time integration was obtained with the second order middle point scheme with the timestep $\Delta_\tau = 0.02$. The values of the model parameters $\alpha = \frac{A}{\Omega_E}, \mu = \frac{M}{\Omega_E R_E^2}$ and $k_H = \frac{\sqrt{2}H_0}{H_A}$ are given in Table 1.

The orography field $h$ (Fig. 1b) is constructed by projecting the real 1° resolution orography onto the T21 antisymmetric spherical harmonic basis. Finally, we construct the external forcing following the procedure proposed in [62]. Let's first observe that the time-averaging of (12) under conditions of steady state allows us to write $\overline{f_{ext}}$ as

$$\overline{f_{ext}} = (\nabla^2)^{-1}\overline{J(\psi, \nabla^2\psi + l + k_H h)} + \alpha\overline{\psi} - \mu\nabla^2\overline{\psi}, \qquad (13)$$

where overbar indicates time averaging. The idea is to use actual atmospheric data in the form of 40 yrs, winter-only averages of the NCEP/NCAR reanalysis data[63] for the streamfunction at the 300mb level in order to calculate the averages on the right hand side of the above formula. The resulting field $f_{ext}$ (Fig. 1a, dimensionless units) is then used as forcing in the model given in Eq. (12). This provides a rough yet effective way to force our simple atmospheric system to resemble the real winter atmosphere of the northern hemisphere.

| Parameter/variable | Symbol | Value | Scaling factor | Nondimensional Value |
|---|---|---|---|---|
| Earth radius | $R_E$ | 6.37x10$^6$ m | - | 1 |
| Earth angular velocity | $\Omega_E$ | $2\pi$ day$^{-1}$ | - | 1 |
| Height of the atmosphere | $H_A$ | 10$^4$ m | - | - |
| Orography normalization | $H_o$ | 10$^3$ m | - | - |
| Coriolis parameter | $L = 2\Omega_E \sin\vartheta$ | - | $\Omega_E$ | $l = 2\sin\vartheta$ |
| Orography | $H$ | - | $H_o$ | $h$ |
| External forcing | $F_{ext}$ | - | $\Omega_E$ | $f_{ext}$ |
| Stream function | $\Psi$ | - | $\Omega_E R_E^2$ | $\psi$ |
| Eckman friction | $A$ | 0.04 day$^{-1}$ | $\Omega_E$ | $\alpha = 6.3 \cdot 10^{-3}$ |
| Turbulent diffusion | $M$ | 1.785x10$^{10}$ m$^2$ day$^{-1}$ | $\Omega_E R_E^2$ | $\mu = 7.0 \cdot 10^{-5}$ |
| Orographic coefficient | $K_H = \sqrt{2}\Omega_E/H_A$ | 8.9x10$^{-4}$ m$^{-1}$ day$^{-1}$ | $\Omega_E/H_o$ | $k_H = 0.1414$ |

Table 1. Variables and parameters of the model. 1 day = 8.64x10$^4$ s.

With this particular choice of parameters, the model is able to simulate fairly well the average state and variability of large scale atmospheric circulation. On Fig. 2 (left column) one can see the mean and variance derived from 40 winters of the NCEP/NCAR data for the 300 mb streamfunction. Note that we plot the fields using dimensionless units obtained as a result of applying a normalization through the factor $R_E^2/\Omega_E$. The right column of Fig. 2 portrays the same statistical properties, yet produced with the T21 model given in Eq. (12). The average state is reproduced rather well, as (partly) expected from the way we have constructed the forcing. The variability in the model is weaker than in the real data, as a result of neglecting all explicit baroclinic processes, but its spatial structure is approximately correct.

Using the standard approach proposed by Benettin et al. [64] (see recent review in [55]) we have calculated the Lyapunov spectrum of the model. We find 28 positive Lyapunov exponents (Fig. 3a). The charateristic e-growth time of the largest Lyapunov exponent $\lambda_1$ is 6 days, which is in rough agreements with the expected predictablity of the large scale weather patterns at 300 mb, while the Kolmogorov-Sinai entropy $h_{KS}$, given by the sum of the all positive Lyupunov exponents, has a value of about 2 days$^{-1}$. Additionally, the Kaplan-Yorke dimension $d_{KY}$ is about 65.7. Note that we have chosen a rather weak boundary layer friction in the model (compared to, e.g., what used in [40]) in order to guarantee strong chaoticity in the system: this is crucial for being on the safe side when taking into account the Gallavotti-Cohen chaotic hypothesis[19].

In Fig. 3b we plot the power spectra of the time series of two relevant global quantities, namely the energy ($\sum_{i=1}^{Nd} -\kappa_i \psi_i^2$) and the enstrophy ($\sum_{i=1}^{Nd} \kappa_i^2 \psi_i^2$) where $\kappa_i$ is the (negative) eigenvalue of the Laplacian operator $\nabla^2$ corresponding to the spherical harmonic $Y_i$ ($\psi(t) = \sum_{i=1}^{Nd} \psi_i(t) Y_i$). We note that such power spectra have no distinct features except a broad maximum in the low frequency range suggesting lack of specific resonances modulating the chaotic behavior on the attractor.

## 4. Results

Our goal is to study the response of the model given in Eq. (12) to changes in its parameters $\alpha, \mu, k_H$ and in the amplitude of the external forcing $f_{ext}$. As a first step, starting from initial conditions of rest, we integrate our model for 10000 days using the reference values of the parameters in order to converge to the attractor.

The most straightforward way to study the model's response is via an ensemble approach. We then integrate our model for additional 3000000 days and chose 15000 equally distant states on this trajectory. Such a 15000-member ensemble on the attractor of the model will be used in all of our calculations and represents our sampling of the unperturbed measure. Several additional experiments with twice as large ensemble have shown that all reported results are stable with respect to the ensemble size. We note that the properties of the time-dependent perturbed system can be described using the concept of pullback attractor a time dependent geometrical object supporting the time-dependent physical measure of the system [65,66,67]. See [35] for a discussion of the link between the theory of pullback attractor and response theory.

Let us apply a perturbation $\pi \to \pi' = \pi + c(t)\Delta\pi$ to a parameter $\pi$ of the system in Eq. (12), where $c(t)$ defines the time protocol and $\Delta\pi$ measures the size of the change. The other parameters are, instead, kept fixed. The resulting change in the expectation value of the streamfunction of the system is defined as $\Delta\langle\psi\rangle_{c,\pi+\Delta\pi}(t) = \langle\psi\rangle_{c,\pi+\Delta\pi}(t) - \langle\psi\rangle_0$. Here $\langle\cdot\rangle_{c,\pi+\Delta\pi}(t)$ is the average over the ensemble of states for the perturbed system, with reference to the changed parameter $\pi$, while $\langle\cdot\rangle_0$ is the average unperturbed system.

In order to compute $\Delta\langle\psi\rangle_{c,\pi+\Delta\pi}(t)$ we construct 15000 ensemble members where the system is perturbed according to $\pi \to \pi + c(t)\Delta\pi$. In our study we will mainly focus in the linear part of the response so that we analyze the following:

$$\langle\psi\rangle^{(1)}_{c,\pi,\Delta\pi}(t) \approx \frac{\langle\psi\rangle_{c,\pi+\Delta\pi}(t) - \langle\psi\rangle_{c,\pi-\Delta\pi}(t)}{2} \qquad (14)$$

As a result, for every $c(t)$ we run two ensembles of perturbed system (for positive and negative perturbations $\pm\Delta\pi$ and same protocol $c(t)$) and estimate the linear component of the response, also by taking into consideration different values of the perturbation strength $\Delta\pi$. Of course, the same procedure could be applied to estimate the response of any observable of the system.

Next, we will discuss the results obtained for the system perturbed by the changes of $\alpha$ and $k_H$, so that $\pi = \alpha, k_H$, thus studying the effect of perturbations to the boundary layer friction and to the orography forcing (a change $k_H \to k_H + c(t)\Delta k_H$ corresponds to a time-dependent rescaling for the orography of the form $h \to h + c(t)h\Delta k_H/k_H$). The reason is that, in general, the effects produced by the changes in $\mu$ are similar to that of in $\alpha$ and the same is true for $k_H$ and amplitude of $f_{ext}$. It should be pointed out that perturbing either $\alpha$ or $k_H$ does not correspond to a simple additive forcing, because it produces additional terms $-(\Delta\alpha)\psi$ or $(\nabla^2)^{-1}J(\psi,(\Delta k_H)h)$ on the right hand side of Eq. (12), respectively, which have an explicit dependence on the state variable. The average effect of the perturbation to the boundary layer friction (orography) is proportional to $-\langle\psi\rangle_0$ ($\langle(\nabla^2)^{-1}J(\psi,h)\rangle_0$), and is shown on Fig. 4a (Fig. 4b).

**4.1. Extracting the Linear Component of the Response**

Following Eq. (14), a first important issue is to test how linear is the system response to the changes of its parameters and to learn how to extract the linear component of the response from the actual response to a finite-size perturbation. We show results obtained for perturbations applied to the orographic forcing ($\pi = k_H$), but similar conclusions hold for the case of boundary later friction ($\pi = \alpha$),

The results of the linearity test for perturbations applied to the orography ($\pi = k_H$) are shown on the Fig. 5. In the left column we show the ensemble averaged system response at days 18, 18.8, 19.4, 20 and 20.6 of the streamfunction field to the pulse (acting only one model time step $\Delta\tau$) of the form $\Delta k_H(t) = 0.3\delta(t)k_H/\Delta_\tau$, where $\delta(t)$ is the Dirac delta distribution; the plotted response is divided by 0.3 in order to scale it with the intensity of the forcing. The system's response to a forcing having the same pattern as before but weaker amplitude ($\Delta k_H(t) = 0.2\delta(t)k_H/\Delta_\tau$) is shown in the central column of the Fig. 5, where in this case, consistently, the field is divided by 0.2. We have also perturbed the system using the protocol $\Delta k_H(t) = 0.2\Theta(t)k_H$. As it follows from the Green's function formalism, the ensemble averaged linear response to a $\delta(t)$ forcing is equal to the time derivative of the system response to a $\Theta(t)$ protocol forcing $(d(.)(t)/dt \cong ((.)(t + \Delta_\tau) - (.)(t))/\Delta_\tau)$. The result of such a calculation is shown in the right column on Fig. 5, where the time derivative of the response to the $\Delta k_H(t) = 0.2\Theta(t)k_H$ time protocol is divided by 0.2 for consistency with the two previous cases.

Comparing these results one can see that the estimate of the linear part of the ensemble averaged response is rather robust, as different ways to evaluate it provide almost indistinguishable results, either by considering forcings of different magnitude or different protocols of forcings. The most relevant property of the linear response of the model shown in Fig. 5 is the presence of a clear quasi-periodic feature with characteristic time scale of 2.6 days. The reasons for this behavior will be analyzed later.

It is important to mention that the linear terms are obtained for relatively strong perturbations (e.g. 30% change of the orography), and that we estimate the nonlinear response of the system as $(\langle\psi\rangle_{c,k_H+\Delta k_H}(t) + \langle\psi\rangle_{c,k_H-\Delta k_H}(t))/2 - \langle\psi\rangle_0$, which is, in some of the analyzed cases, almost twice as large than the linear one. This fact indicates that it is possible to efficiently extract information of the linear response also when strong perturbations are applied. The same conclusions hold if we change other system's parameters.

**4.2. Geometry of the Forcings**

As mentioned above, it is crucial to understand the geometrical relationship between the applied perturbations and the tangent space of the attractor of the system at any given point along the trajectory of the unperturbed system. The best way to address this issue is by calculating the CLVs of the system and estimating the angles between the perturbation and CLVs at each point of the attractor. When the perturbation is parallel to CLVs corresponding to the positive Lyapunov exponents, the system is excited along the smooth part of the attractor, because the invariant measure is absolutely continuous along the unstable manifold, see [16] for details. Perturbations along CLVs corresponding to the negative Lyapunov exponents push the system along the stable direction, where the fractal structure of the attractor is apparent[16]. We know that the natural fluctuations of the system never explore, by definition, the stable directions. Therefore, one expects that occurrence of *climatic surprises*, in the form of occurrence of forced variability that is almost absent in the unperturbed system might result from perturbations having a substantial projection along the stable directions. This is a situation where the FDT could be ineffective in predicting the response to perturbations from the knowledge of the fluctuations of the unperturbed system.

We have constructed the CLVs using the Kuptsov and Parliz procedure[68]. We have tested the property of the CLV covariance (Eq.3) and have found it to be valid with extremely high accuracy (order of $10^{-12}$ over a period of 50 days). Similarly, the corresponding Lyapunov exponents are well separated, suggesting that the algorithm has converged well.

Next, we study the spatial correlation between each CLV and the perturbation vector in the two considered cases (i.e. $-(\Delta\alpha)\psi$ for perturbations to $\alpha$ and $(\nabla^2)^{-1}J(\psi,(\Delta k_H)h)$ for perturbations to $k_H$) at every point of 10000-day trajectory (part of the 3000000-day trajectory mentioned previously).

Figure 6 shows the time averaged fraction of the cumulative norm of the applied forcing projecting on the CLVs of the system, where the CLVs are ordered from the most unstable to the most stable. The first 28 CLVs correspond to the unstable directions, one corresponds to the neutral direction of the flow, and the remaining 202 correspond to the stable directions (see Fig. 3a),

One finds that in the case of perturbations applied to the boundary layer friction the unstable directions give the overwhelming contribution to the variance (over 90%). Since the change in the boundary layer friction results into a forcing that is proportional to the streamfunction itself, it is hardly unexpected that the span of the unstable CLVs can explain most of its variance, see also discussion in [29,30].

Our findings are indeed different when considering the orographic forcing, where the perturbation projects substantially also on the stable directions of the flow. We can then expect that the response of the system to perturbations to the orography might have unexpected features, as discussed in detail below.

**4.3. Time and Frequency Dependent Linear Response**

We first analyze the properties of the system's response to changes in the orography determined by the perturbation $k_H \to k_H + 0.3\delta(t)k_H/\Delta_\tau$. We investigate the Green's function $G^{(1)}_{\Phi,k_H}(t)$ and the corresponding susceptibility $\chi^{(1)}_{\Phi,k_H}(\nu)$, where the observables $\Phi$ we consider are the kinetic energy and the enstrophy. We see that in both cases the Green's function shows pronounced oscillations with the period of about 2.6 days; see Fig. 7a and 7b, top frame.

Let's try to better understand these quasi-regular fluctuations of the Green's function by looking at the susceptibility. We find a clear evidence for the resonant response hinted at above by looking at the susceptibility for the energy and the enstrophy (Fig. 7a and 7b, middle frame) where one can see distinct features for frequencies around 0.4 day$^{-1}$. The quality of the data analysis is strongly supported by the fact that the real and imaginary parts of the susceptibility obey closely the Kramers-Kronig relations corresponding to Eq. (8).

Most interestingly, the behaviour of the unperturbed model is qualitatively different from the scenario given above, as no feature appears at 0.4 day$^{-1}$ in the power spectrum of either the energy or the enstrophy (see Fig. 3b). We will explore in a later section the origin of the resonant response.

We can learn an additional interesting piece of information by looking at the suceptibility. The static susceptibility (zero frequency response) is negative for the energy, implying that raising the orography leads to a decrease of the energy as a long-term effect, while the opposite holds for the enstrophy. This means that higher spatial frequency modes of variability become disproportionally populated as a result of an increase of the orography.

A substantially different system behavior is found when the system is perturbed by changing the coefficient of the boundary layer friction, see Fig. 8. There is no manifestation of the new resonance found when changing the orography of the system. The Green's functions for energy and enstrophy (top frames in Fig. 8a and Fig. 8b, respectively) decay following approximately an exponential behaviour, and, correspondingly, the susceptibilities feature a resonance corresponding to planetary wave activity (time scale of about 20 days), which is present also in the natural variability of the system, see Fig. 3b and the middle and lower frames in Fig. 8a and Fig. 8b. We wish to remark that, as in the case of Fig. 7a and Fig. 7b, the Kramers-Kronig analysis confirms that the produced data are extremely accurate.

Looking at the static susceptibilities, we find, unsurprisingly, negative values for both the energy and enstrophy observables, which implies that increasing the Ekman boundary layer coefficient leads to a decrease for both energy and enstrophy.

**4.4. Skill of the Fluctuation-Dissipation Theorem in Reproducing the Response**

The presence of a stronger projection of the forcing along the stable manifold in the case of orographic forcing vs. in the case of changes in the boundary layer friction suggests that in the former case the skill of the FDT in reconstructing the response should be worse. Our unperturbed model lives in a regime of strong chaos, so that, following what observed in [69], one expects that the probability distribution function (PDF) of large-scale climate observables be to a good approximation Gaussian. Therefore, it makes sense to use the simple quasi-Gaussian approximation when applying the FDT (qG-FDT). This amounts to constructing by best fit of the first and second moments an approximate Gaussian model for the invariant measure, and then using the resulting fitted distribution in Eq. (9). The estimates obtained via FDT of the Green's functions describing the impact of changing the coefficients $\alpha$ and $k_H$ for the energy and enstrophy observables are shown in Fig. 10. One finds that in all cases there is a considerable difference between the true Green functions (see also Fig. 7a-b and Fig. 8a-b) and those derived using the FDT, both on short and on long time scales. The agreement is clearly worse in the case of the orographic forcing, as anticipated, where the FDT-reconstructed Green's function is only qualitatively similar to the true one. Interestingly, if one applies a truncation to the FDT-reconstructed Green's function obtained through projection on the leading EOFs on the unperturbed system – see [70] for a detailed explanation of the procedure - the estimate of the Green's functions is substantially improved. This suggests that the FDT provides an imprecise estimate exactly of those contributions to the Green's function coming from marginal variability modes of the system. This makes sense, as these might be though to be disproportionally represented by the stable modes of the flow. The optimal performance is obtained using a truncation including 200 EOFs for the case of perturbations to the orographic forcing and 90 EOFs in the case of perturbations to the boundary layer friction. Note that, obviously, the truncated forcings look very similar to original ones, but the qG-FDT is, instead, sensitive to the dimension of EOF basis, the reason being that the EOF truncation gets rid effectively of most of the (wrongly estimated) contributions coming from some of the stable directions.

**4.5 Resonant Response and Unstable Periodic Orbits**

Another way to describe the response of the system is by looking at its leading variability patterns. the analysis of the variability is through the use of the empirical orthogonal functions (EOF) or complex EOF (CEOF) decomposition. EOFs are by definition the eigenvectors of the covariance matrix of the time series of the variables of the system, so that the leading EOF is the pattern describing the largest portion of the overall variabilty of the data. The leading CEOF can be used to construct a 2D plane in the phase space where the field of interest has maximum rotational variability (see [71] for definitions and [72] for application example). By comparing the leading CEOFs of the response fields with those of the unperturbed model runs one can see whether the response of the model has different structure with respect to the model natural variability.

In the rest of the paper we focus only on the system response to the orographic forcing detailed above. Indeed, as demonstrated by the top frame of Fig. 9, the leading CEOFs of the response (capturing almost 90% of the variability in a 1-7 day interval) has large scale (wave number 2) structure. Further analysis shows that it propagates eastward and could be identified as an orographically forced Rossby-like planetary wave. In the same time-scale range, the two leading CEOFs of the model (Fig. 9, bottom and middle frames) explaining 40% of the natural variability are small scale (mainly projecting on wave numbers 6 to 7) patterns moving from east to west. They could be associated with small scale Rossby waves trapped in the subtropical jet. Clearly, the two features are entirely different.

In Fig. 11 we show the time evolution of the PDF anomaly for the streamfunction of the model forced by the perturbation in the orography. We portray its projection onto the CEOF1 of the response shown in the top panel of Fig. 9. We have estimated the PDF projections using standard histograms. We have constructed 20x20 equispaced (along the x- and y-directions) bins and computed the occupation numbers (from the ensemble members) for both perturbed and unperturbed ensembles. The normalized differences between the occupations numbers (i.e. the PDF anomaly) realized in the forced runs and in the control run at times 0.4, 0.8, 1.2, 1.6, 2.0, 2.4, 2.8, 8.0, 13.2, 18.4, 23.6, 28.8 days are shown in Fig. 10. The result is robust with respect to the way we choose the bins. From Fig. 11 one can see an additional indication of the resonant response of the system. The PDF of the perturbed system rotates in the CEOF1 plane around the unperturbed PDF of the system with a period of approximately 2.6 days, while its amplitude decays slowly to zero on a time scale of about 20 days, which corresponds to the small spectral width of the resonance peak shown in Fig. 7.

In order to to understand the resonant response behaviour, it is not possible, as clear from the discussion above, to use the usual argument of a forcing amplifying some prominent feature of the natural variability of the system. We have then studied the set of the UPOs of the system calculated in [41]. This set contains a cluster of UPOs having a period close to the 2.6 days. These orbits are oriented in the phase space in a direction almost parallel to the CEOF1 of the response. All of the UPOs from this cluster are much more unstable than the typical system trajectory, since they have 40-50 unstable directions, in contrast with the 28 positive LEs evaluated on the system's attractor, and hence play an absolutey marginal role in determining the natural variability of the system. The projections of these orbits on the CEOF1 of the response are shown on Fig. 12 together with the projection of an unperturbed trajectory. One can see that these UPOs do not belong to the core of the attractor of the system and has visual confirmation of the fact that their neighborhoods are only very rarely visited by the system. On the other hand, the orographic forcing acts substantially along the stable directions of the attractor and pushes the PDF of the system along the region of the phase space dominated by the UPOs described above. When the forcing is halted, the PDF of the system PDF rotates along these orbits and relaxes progressively towards the its unperturbed position. Summarizing, the cluster of UPOs described here nicely explains why

a) orographic forcing has a relatively low projection on the unstable directions of the flow;

b) why it provides an unexpected resonant response; and

c) why the natural and forced variability of the system have in this case little relation, thus showing a violation of FDT-kind arguments.

The resonant behavior obtained in a case of the orographic forcing could be qualitatively explained by the linear Rossby theory, because one of the complex eigenvectors of the linearized operator of the model around its average state has a structure very similar to that of the resonant response pattern (top frames of Fig. 9). On the other hand, the linear Rossby theory overestimates the magnitude of the response by a factor of 3 and cannot predict the existence of regular oscillations in the response after 5 days. This is shown on Figs. 10 b) and d), taking into account that the qG-FDT holds exactly for the linearized model.

## 5. Summary and Conclusions

The main aim of this paper has been to study the response of a quasi-geostrophic barotropic system tuned to represent the winter 300 mb atmosphere of the northern hemisphere to perturbations and to provide a dynamical interpretation of the obtained results. This exercise has relevance within the overall goal of improving our understanding of how nonequilibrium statistical mechanical systems, and, in particular, chaotic geophysical flows, respond to perturbations. The conceptual framework for our analysis comes from adopting the chaotic hypothesis by Gallavotti and Cohen, so that we use mathematical tools borrowed from the theory of Axiom A dynamical systems.

The Ruelle response theory provides us with a clear protocol on how to perform perturbation experiments and derive the linear response for any observable of interest. We focus on two families of experiments, performed by changing the value of the boundary layer friction and the scaling height of the orography, and look at the properties of the average energy and average enstrophy, thus investigating two relevant squared norms of clear physical and meteorological relevance. Using for each forcing an ensemble comprising of 15000 model runs, we derive accurate estimates of the linear Green functions describing the response of the system for each considered observable. We note that we have been able to extract information on the linear contribution to the response even if we have applied not-so-weak forcings, through a simple yet effective algebraic combination of the data giving the response of the system to positive and negative forcings of different strength. We have confirmed the quality of the obtained linear terms by testing that the susceptibilities corresponding to the Green's functions obey the Kramers-Kronig relations to a high degree of accuracy.

The most interesting result we find is that, when considering perturbations to the orographic forcing, we discover an intense and surprising resonant feature corresponding to westward propagating waves with period of about 2.6 days. Such a feature is virtually absent from the natural variability of the unperturbed system: this raises a flag regarding the applicability of the FDT. In fact, we discover that the forcing projects significantly on the stable directions of the flow, as determined through a geometrical analysis performed with CLVs. This is in excellent agreement with the Ruelle theory, which envisages the lack of clear correspondence between natural and forced fluctuations exactly in this dynamical scenario. We have verified that the quality of the FDT-based reconstruction of the Green's function is better (even if far from perfect) in the case of perturbations to the value of the boundary layer friction, while, as anticipated, the procedure basically fails when considering the change in the orographic forcing.

The observed resonance can be explained using a different reconstruction of the invariant measure of the unperturbed system. We find that a subset of the system UPOs characterized by several large positive LEs is responsible for this behavior. Note that the strong instability of these orbits means exactly that their neighborhoods are extremely rarely visited by the unperturbed system. This further clarifies the reason for the lack of correspondence between forced and free fluctuations.

Our results clarify that in a nonlinear chaotic system the information one can derive from the analysis of the natural variability is not enough to predict its response to forcings, as surprises are possible: in other terms, understanding climate variability and predicting climate change are in general two separate, yet related, problems.

We underline that, following[33], we have that the susceptibility of the system can be reconstructed by applying a random forcing and studying the power spectra of the perturbed and of the unperturbed system. What has been found here then suggests that adding a stochastic forcing to a model of climate or weather, as done when implementing stochastic parametrizations[73], might lead to unexpected and extremely relevant (or pathological and spurious) responses, if resonances are excited. We will investigate this problem in future studies.

Some of the tools presented here deserve further investigation with the goal of studying specific meteo-climatic phenomena. In particular, we wish to use the formalism of UPOs for re-examining the classic problem of weather and climate regimes in the context of baroclinic models (as opposed to barotropic models used elsewhere), in the hope of being able to reach a better understanding of the dynamical structures behind blocking events. Taking inspiration from the results shown in [30] we could then be able to explain better the anomalies in predictability and energy processes taking place during blocking events.

**Acknowledgments.** AG received support from Russian Foundation for Basic Research (project 16-55-12015) and Russian Federation government (Agreement 14.Z50.31.0033). AG and VL acknowledge the support received from the DFG Cluster of Excellence CliSAP. VL acknowledges the support received by the DFG project MERCI. VL wishes to express gratitude to M. Ghil for many generous scientific exchanges in the course of these years.

**Figures**

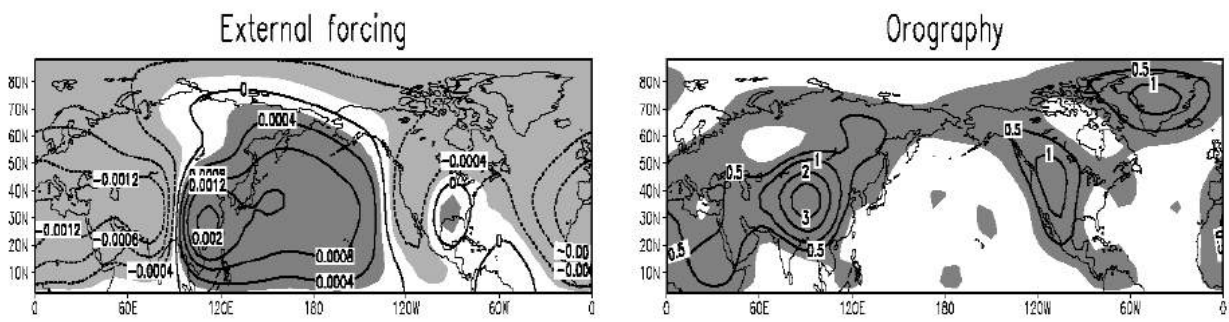

Figure 1. External forcing (dimensionless units) and orography (km) used in the quasi-geostrophic barotropic model of the winter north hemisphere atmosphere.

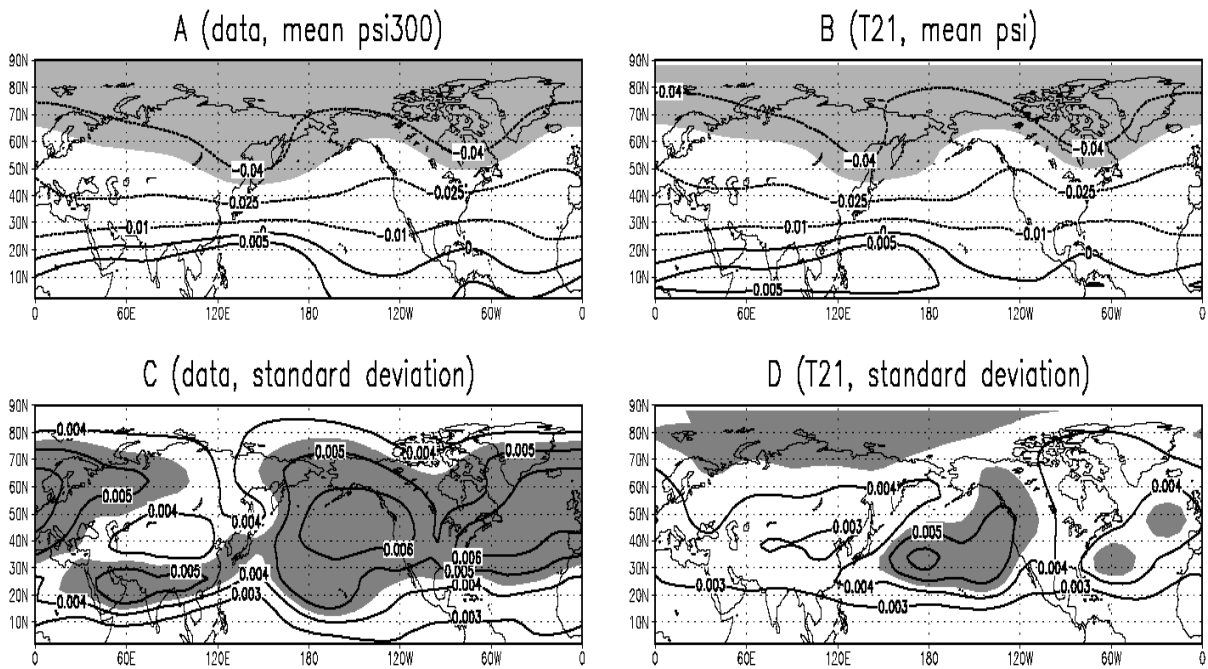

Figure 2. Average streamfunction (B) and standard deviation of the streamfunction (D) of the the quasi-geostrophic barotropic model of the winter north hemisphere atmosphere compared to NCEP/NCAR reanalysis data used to construct the forcing (streamfunction field at 300mb level) ((A) and (C).

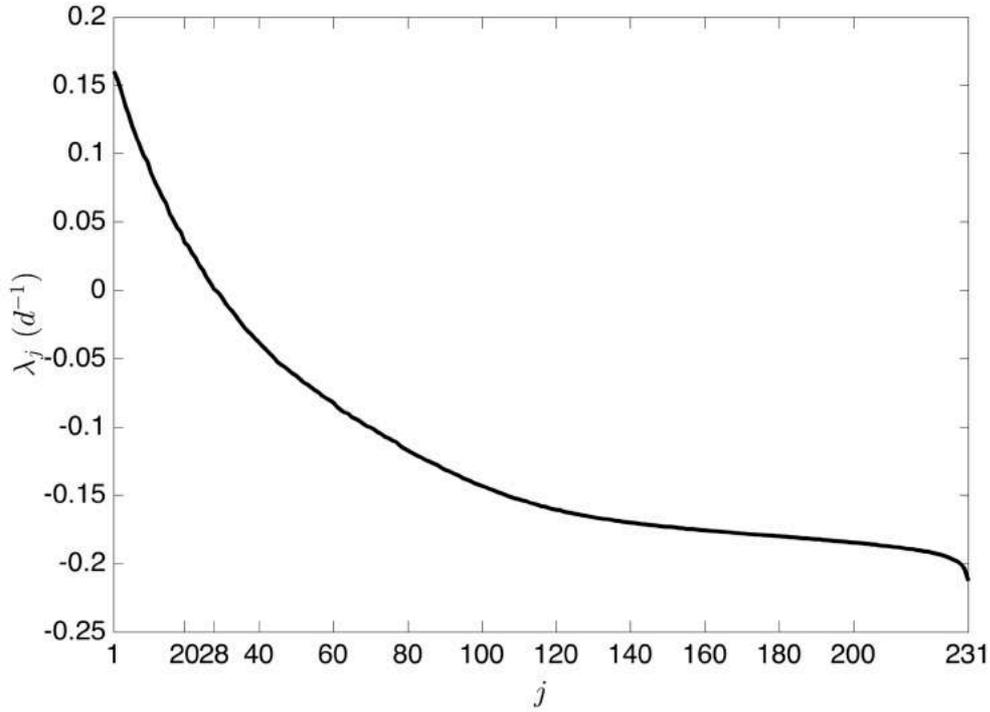

a)

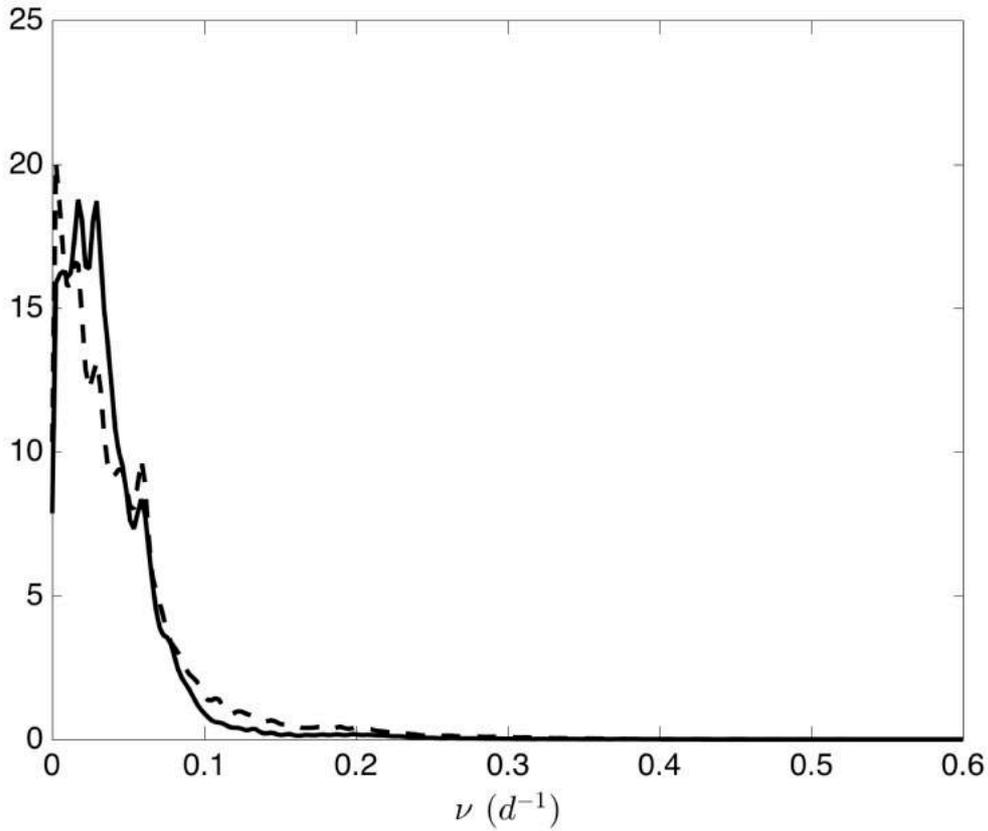

b)

Figure 3. a) Spectrum of the Lyapunov exponents of the unperturbed model. b) Power spectra of the spatial variance of the total energy (solid line), and of the enstrophy (dashed line) of the unperturbed model. The power spectra are normalized with respect to the variance of the corresponding time series, so integrating the curves from 0 to infinity gives 1.

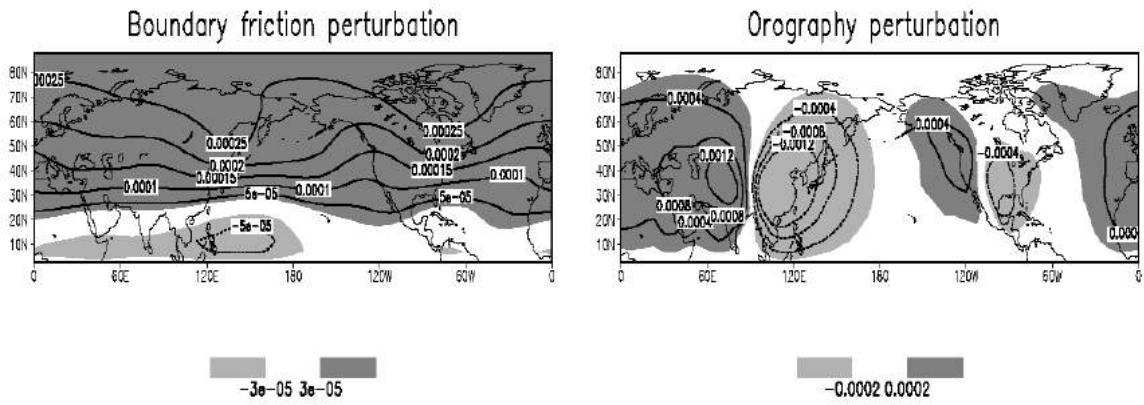

Figure 4. Average effects of perturbations as additional tendencies in Eq.12. Left – change of the Ekman boundary layer coefficient (dimensionless units), right – change of the orography height (dimensionless units).

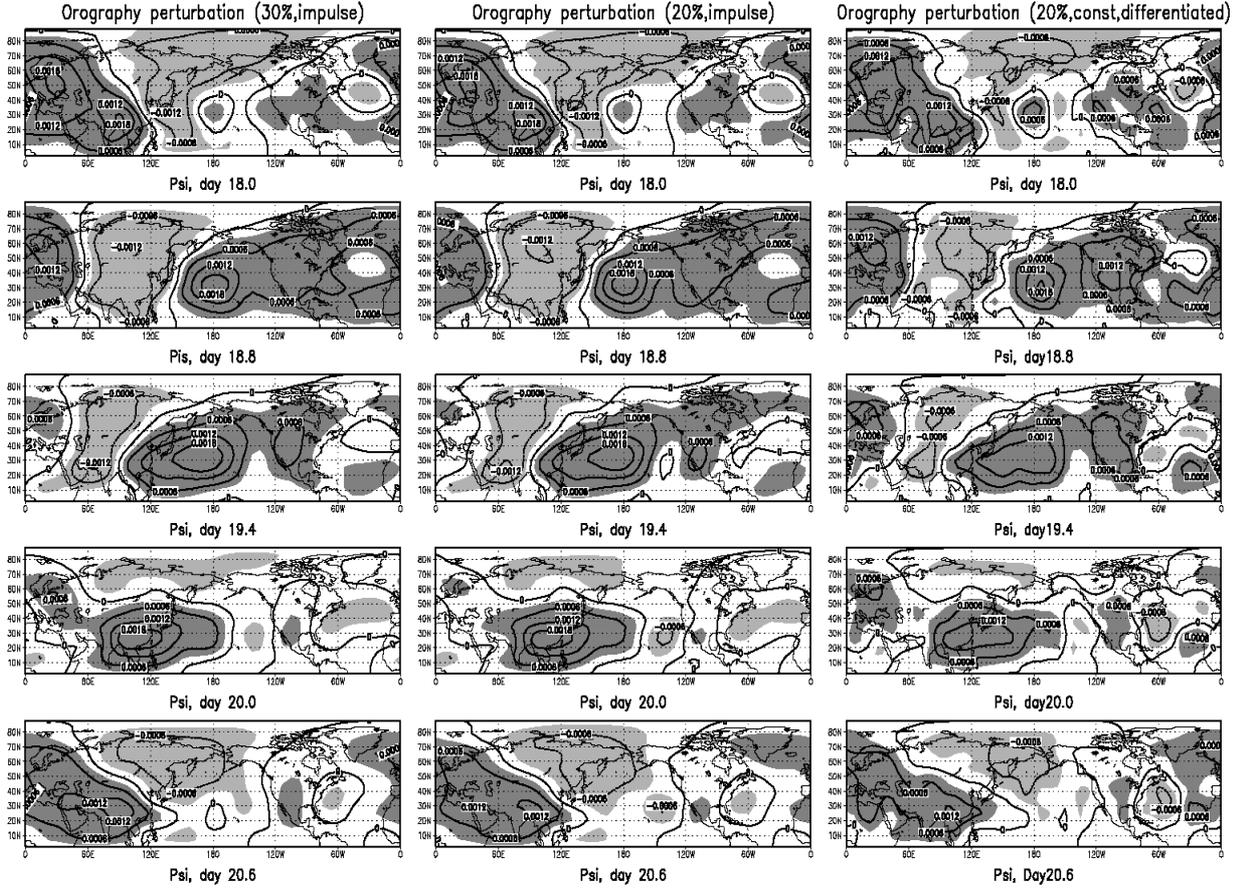

Figure 5. Test of the linearity of the response with respect to the forcing magnitude and scenario. Ensemble averaged model response to perturbation to the orography. Left column: Model response to perturbation to the orography of the form $k_H \to k_H + 0.3 k_H \delta(t)/\Delta_\tau$ where $k_H$ is the background coefficient describing the interaction with orography (responses at days 18, 18.8, 19.4, 20 and 20.6 are shown, divided by a factor 0.3). Central column: same as for the left column one but perturbation is of the form $k_H \to k_H + 0.2 k_H \delta(t)/\Delta_\tau$ (the response is divided a factor 0.2). Right: Reconstructed model response to delta-function forcing normalized as for the left column derived from actual model response to step-like increase of the orography of the form $k_H \to k_H + 0.2\Theta(t)$, where Θ is the Heaviside distribution.

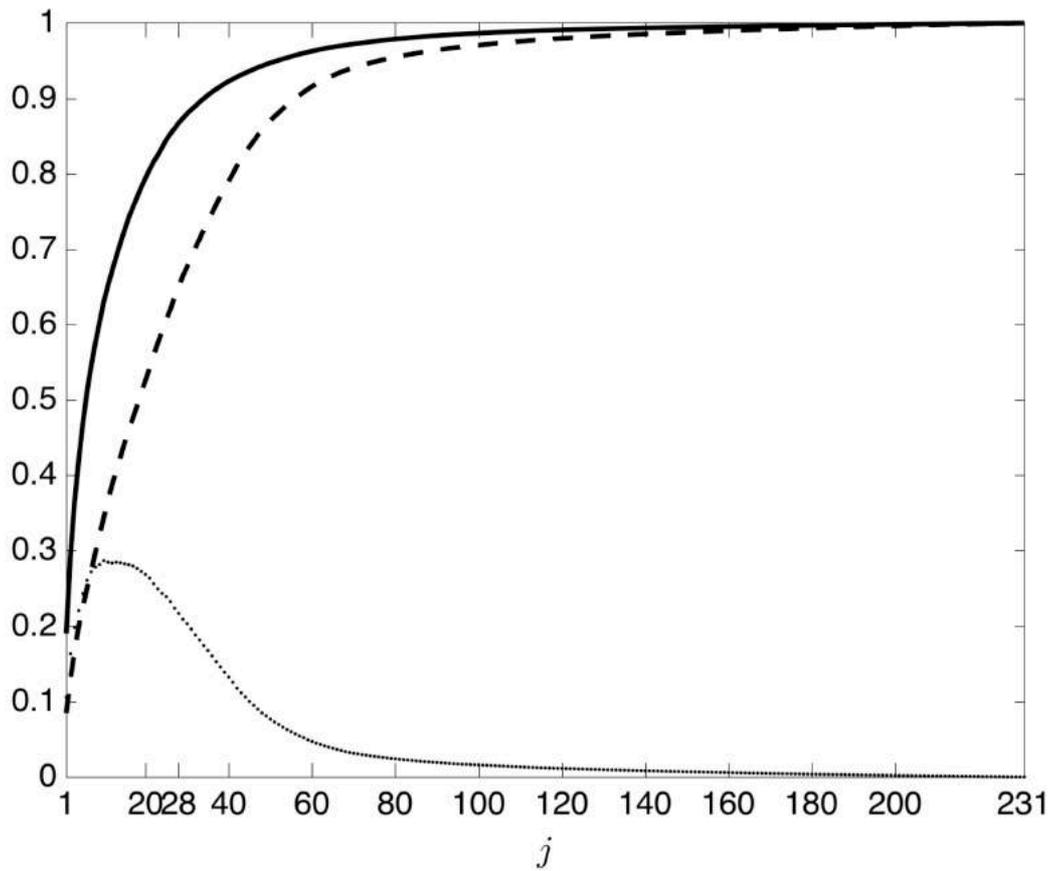

Figure 6. Fraction of the norm of the applied forcing (y-axis) belonging to the subspace of the first $j$ (x-axis) CLVs. Solid line: change in the boundary layer friction. Dashed line: orography forcing. Dotted like: difference between the two curves.

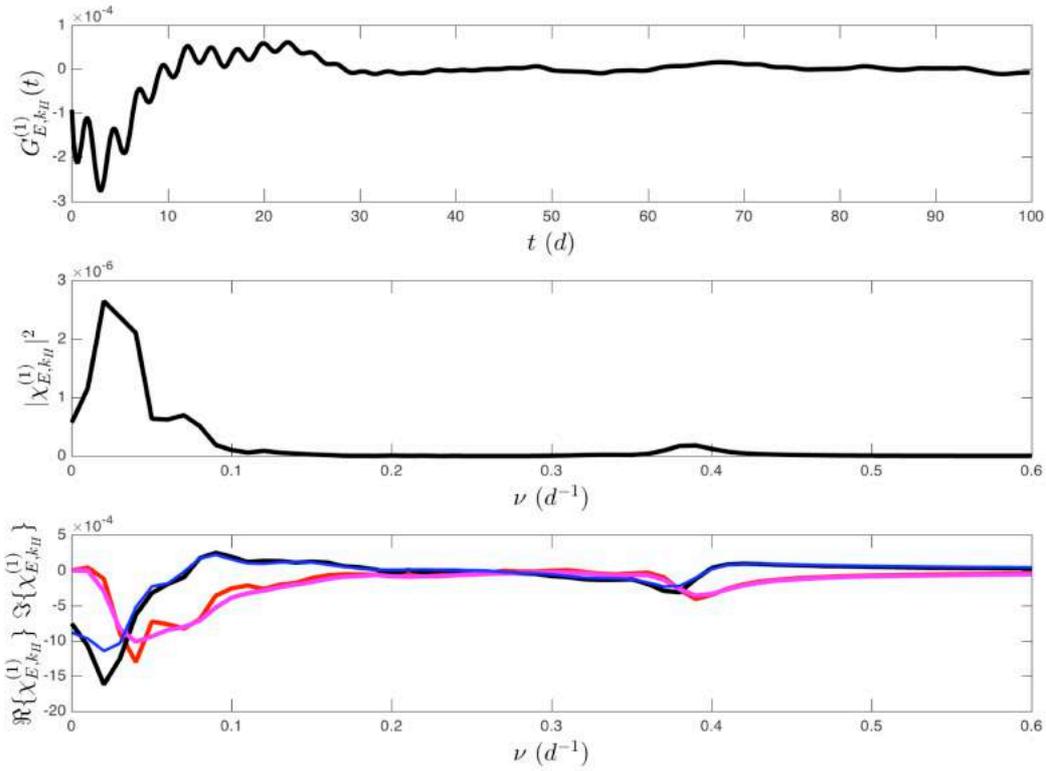

a)

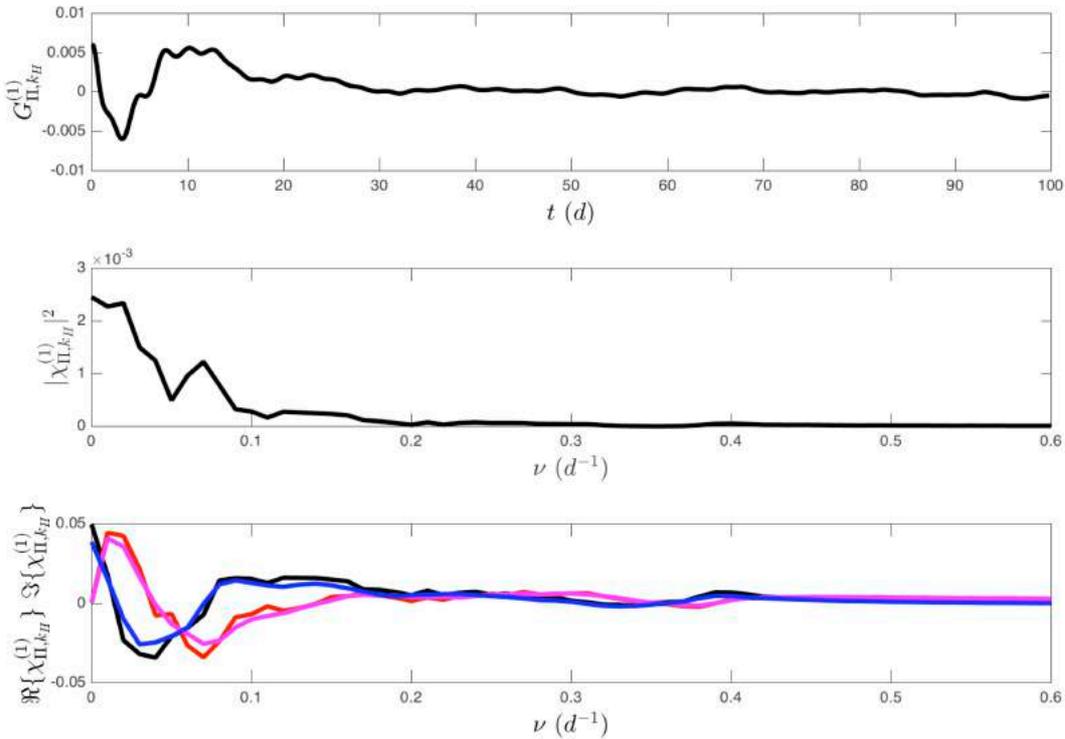

b)

Figure 7. Response of the model to an increase in the orographic forcing. Panel a) Response of the total energy. From top to bottom: Green's function, squared modulus of the susceptibility, and real and imaginary part of the susceptibility. In the bottom graph, the black and the red line show the real and imaginary part of the susceptibility, respectively. The blue and magenta lines show the estimates of the susceptibility obtained using the Kramers-Kronig relations. Panel b) same as a) but for the total enstrophy.

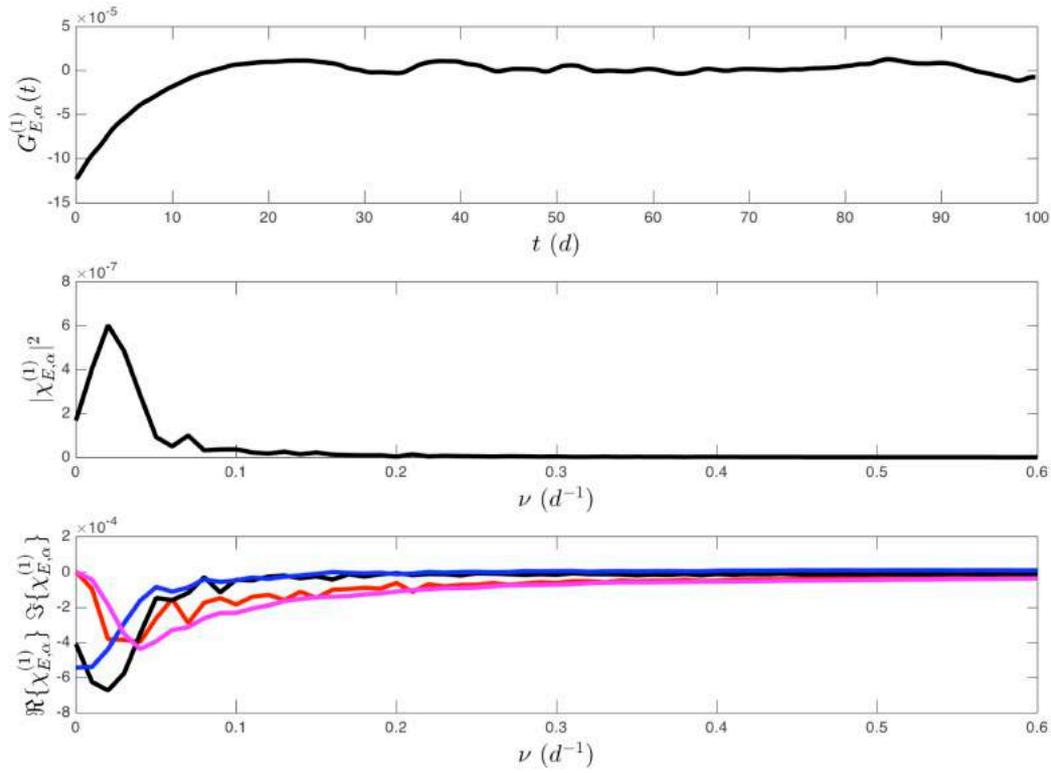

a)

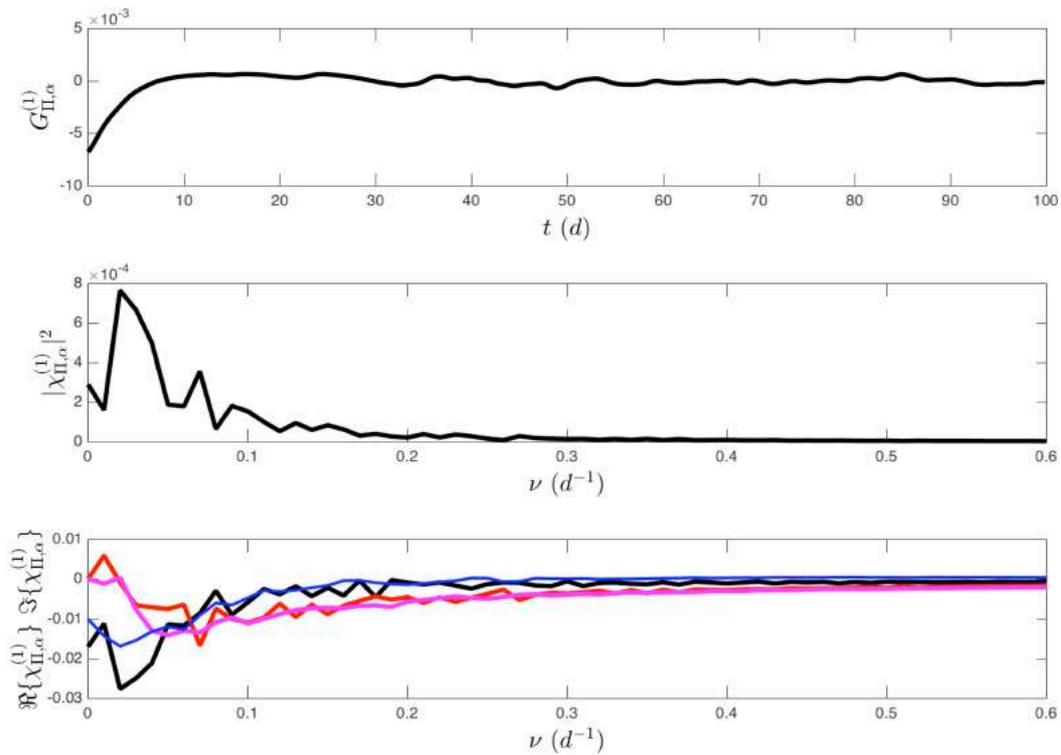

b)

Figure 8. Response of the model to an increase in the boundary layer dissipation. Same as Fig. 8. Panel a): Response of the total energy. Panel b) Response of the total enstrophy.

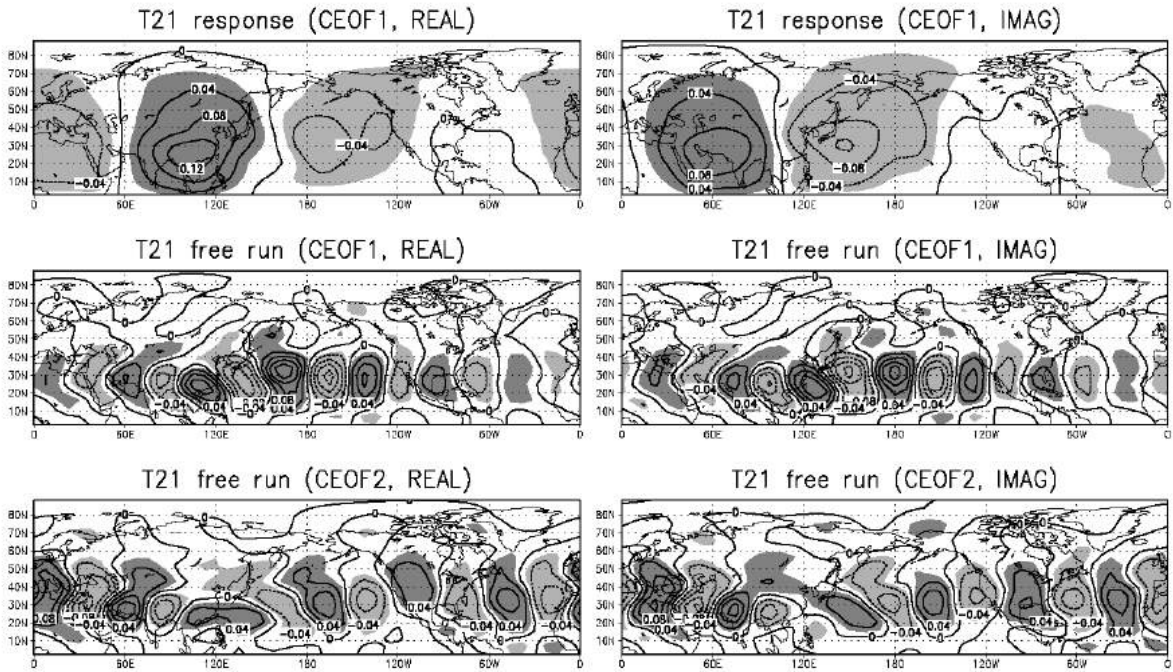

Figure 9. Model response to the orography forcing: CEOF1 (real and imaginary parts) of the system response (upper row). CEOF1 and CEOF2 of unperturbed system (bottom and middle). A 1-7 day bandpass filter has been applied to all data.

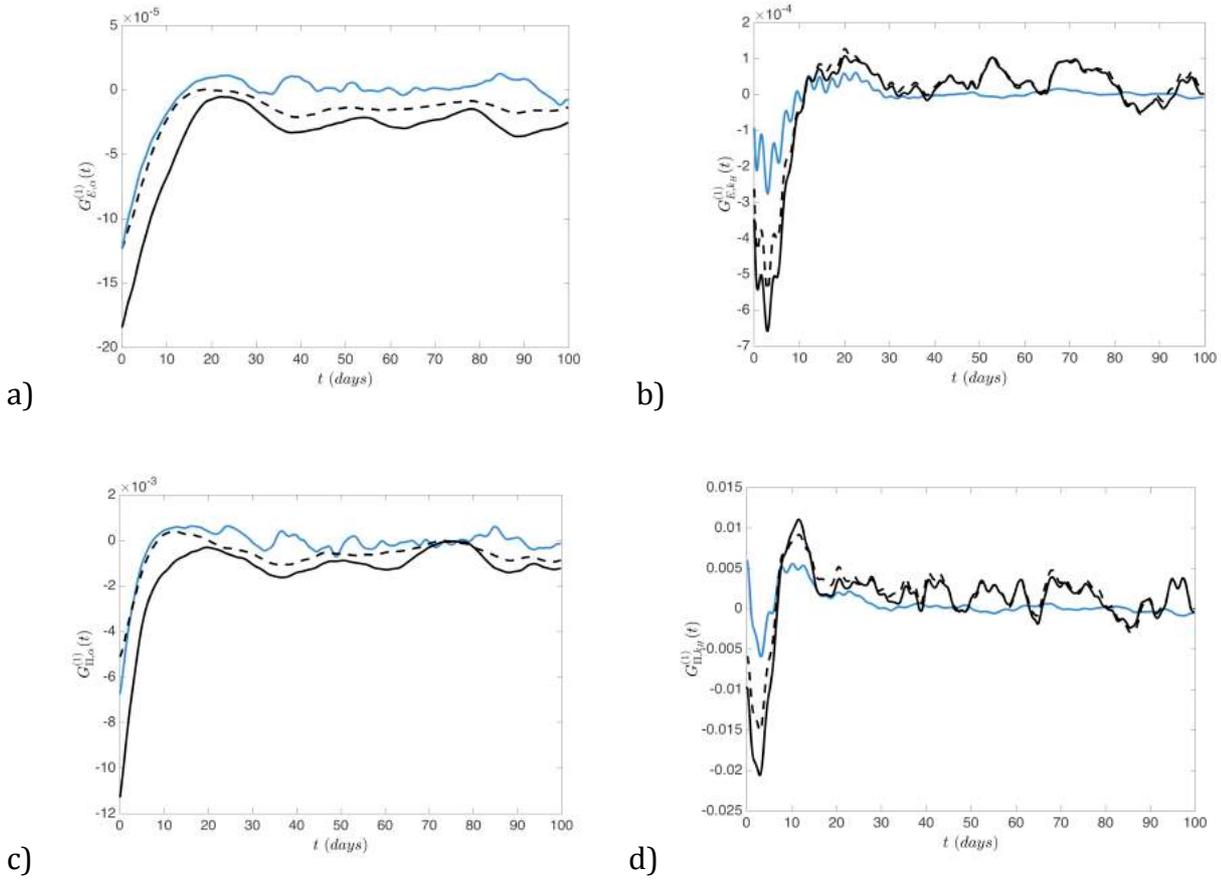

Figure 10. Skill of the FDT in reproducing the response of the system to perturbations. Left panels: response to changes in the boundary layer friction. Right panels: response to changes in the orography. Upper (lower) panels: Results refer to the total energy (enstrophy). The blue lines give the actual response of the system as portrayed in Figs. 7 and 8 (in dimensionless units). The solid black lines refer to the FDT-reconstructed response. The dashed lines refer to the FDT reconstructed response obtained in the reduced space described in the text.

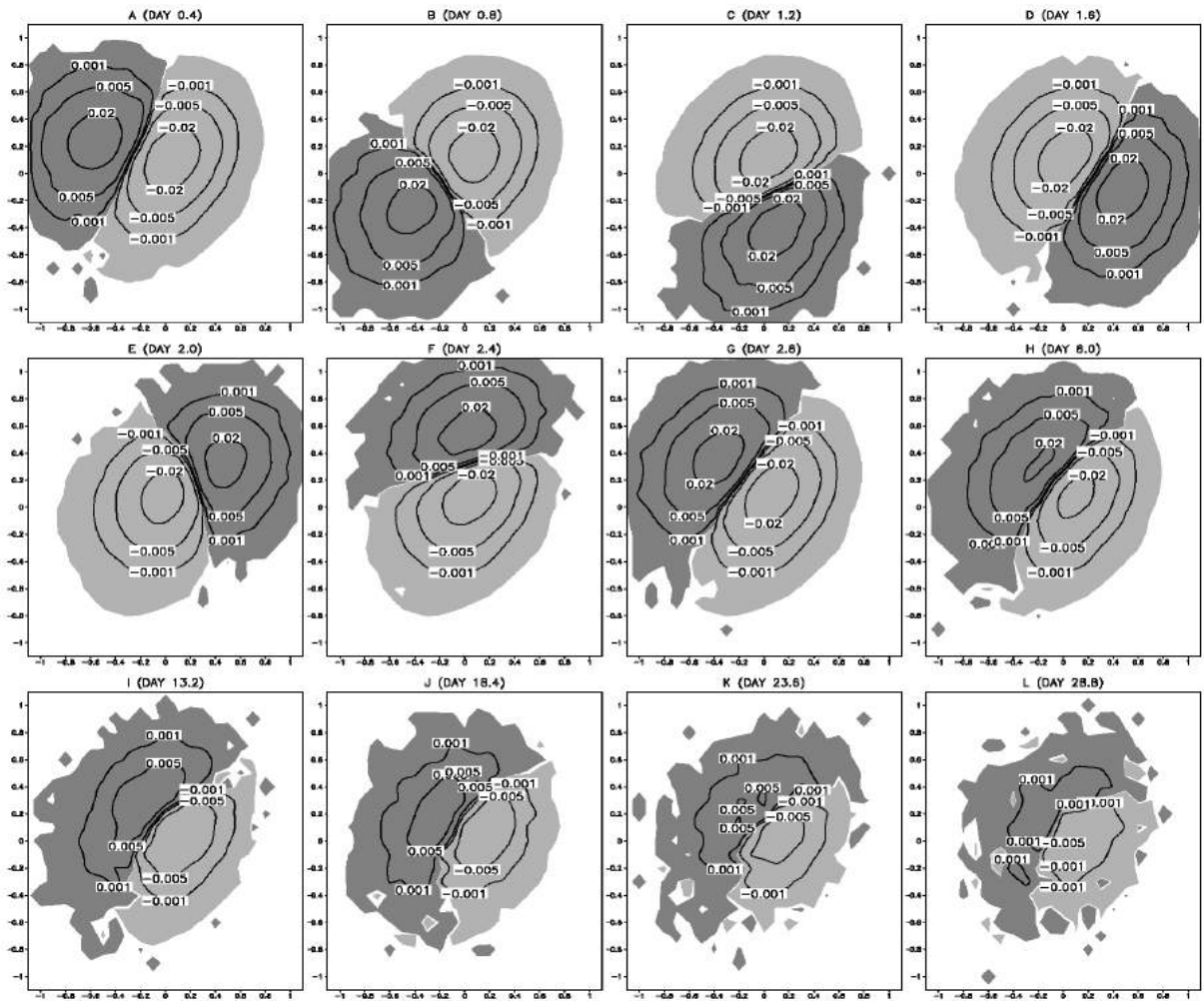

Figure 11. Model response to the orography forcing: time evolution of the PDF anomaly projected on the plane of the CEOF1 of the response (Fig. 9 top frames). X-axis – normalized projection value to Re(CEOF1), y-axis – normalized projection value to Im(CEOF1). Positive and negative values exceeding $3 \cdot 10^{-5}$ in absolute value are colored in dark and light grey. Contour lines are -0.02, -0.005, -0.001, 0.001, 0.005, 0.02.

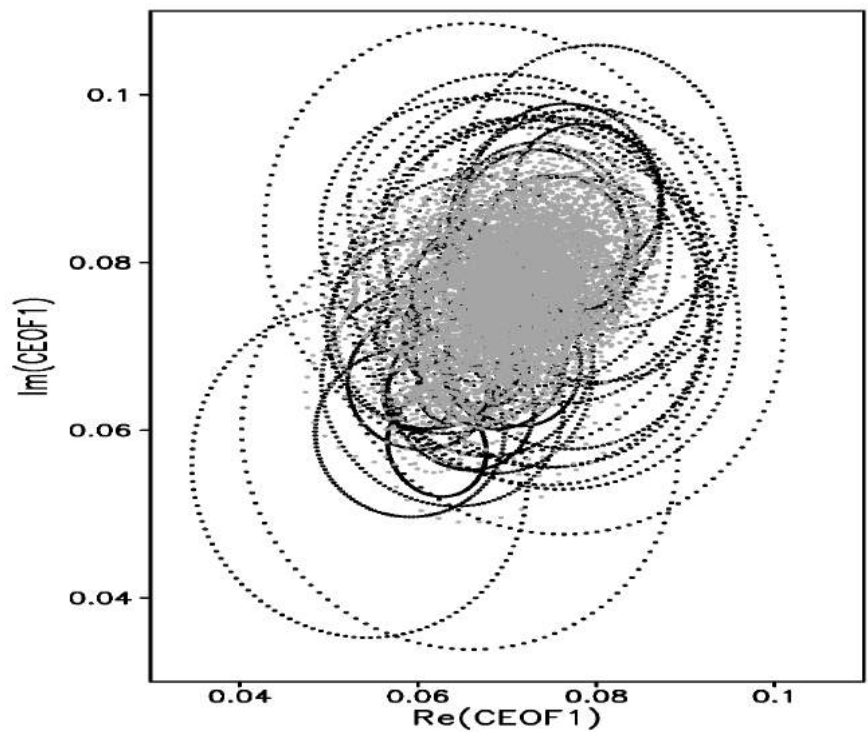

Figure 12. UPOs responsible for the resonant response (black) and model trajectory (grey) projected onto the plane of the CEOF1 of the response (formed by the vectors shown in the top frames of Fig. 9). X-axis – dimensionless projection value to Re(CEOF1), y-axis – dimensionless projection value to Im(CEOF1).

# References

[1] Ruelle, D., Thermodynamic formalism: the Mathematical structures of classical equilibrium statistical mechanics, Addison-Wesley, Reading, 1978

[2] Gallavotti, G., Statistical Mechanics: A Short Treatise Springer, New York, 2013

[3] Gallavotti, G., Nonequilibrium statistical mechanics (stationary): overview, in Encyclopedia of Mathematical Physics 3, 530-539, (J.-P. Franchise, G. L. Naber, T. S. Tsou Eds) Elsevier, 2006

[4] Gallavotti, G., Nonequilibrium and Irreversibility, Springer, New York, 2014

[5] de Groot R. and P. Mazur, Nonequilibrium thermodynamics, North Holland, Amsterdam, 1962

[6] Peixoto, J. P. and A. H. Oort, Physics of Climate, AIP, New York, 1992

[7] Ghil, M., A mathematical theory of climate sensitivity or, how to deal with both anthropogenic forcing and natural variability? In: Chang, C.P., Ghil, M., Latif, M., Wallace, J.M. (eds.) Climate Change: Multidecadal and Beyond, pp. 31–51. Kluwer, Dordrecht, 2015

[8] Lorenz, E. N., The Nature and the Theory of the General Circulation of the Atmosphere, WMO, Geneva, 1967

[9] Lucarini, V., R. Blender, C. Herbert, S. Pascale, F. Ragone, and J. Wouters, Rev. Mathematical and Physical Indeas for Climate Science, Geophys. 52, 809-859, doi: 10.1002/2013RG000446 (2014)

[10] Kubo, K., The Fluctuation-Dissipation Theorem, Reports on Progress in Physics 29, 255-284 (1966)

[11] Marini Bettolo Marconi, U., A. Puglisi, A. Rondoni, and A. Vulpiani, Fluctuation-dissipation: Response theory in statistical physics, Physics Reports 461, 111-195 (2008)

[12] Lucarini,V., and M. Colangeli, Beyond the linear fluctuation-dissipation theorem: the role of causality. J. Stat. Mech. P05013 doi:10.1088/1742-5468/2012/05/P05013 (2012)

[13] Eckmann, J. P., and D. Ruelle, Ergodic theory of chaos and strange attractors, Rev. Mod. Phys. 57, 617-656 (1985)

[14] D. Ruelle, Differentiation of SRB states, Comm. Math. Phys. 187, 227–241 (1997)

[15] D. Ruelle, General linear response formula in statistical mechanics, and the fluctuation-dissipation theorem far from equilibrium, Phys. Letters A 245, 220-224 (1998)

[16] Ruelle, D., A review of linear response theory for general differentiable dynamical systems, Nonlinearity 22, 855-870 (2009)

[17] Lucarini, V., Response Theory for Equilibrium and Non- Equilibrium Statistical Mechanics: Causality and Generalized Kramers-Kronig Relations, J. Stat. Phys. 131, 543-558 (2008)

[18] Lorenz, E. N., Forced and free variations of weather and climate, J. Atmos. Sci. 36 1367-1376 (1979)

[19] Gallavotti, G., and E. G. D. Cohen, Dynamical ensembles in nonequilibrium statistical mechanics, Physical Review Letters 74, 2694-2697 (1995)

[20] Abramov, R. and A. Majda, Blended response algorithms for linear fluctuation-dissipation for complex nonlinear dynamical systems, Nonlinearity 20, 2793 (2007)

[21] Wouters, J. and V. Lucarini, Multi-level dynamical systems: Connecting the Ruelle response theory and the Mori-Zwanzig approach. J Stat Phys, 151, 850 (2013)

[22] Chekroun, M. D., H. Liu, and S. Wang. Stochastic Parameterizing Manifolds and Non-Markovian Reduced Equations,. SpringerBriefs in Mathematics, Springer, NY (2015)

[23] Gritsun, A. and G. Branstator, Climate Response using a three-dimensional operator based on the fluctuation-dissipation theorem, Journal of the Atmospheric Sciences 64, 2558 (2007)

[24] Gritsun, A., G. Branstator, and A. Majda, Climate response of linear and quadratic functionals using the fluctuation dissipation theorem, J. Atmos. Sci. 65, 2824 (2008)

[25] Cooper F. C., and P. H. Haynes, Climate Sensitivity via a Nonparametric Fluctuation-Dissipation Theorem. J. Atmos. Sci. 68 937-953 (2011)

[26] Eyink, G. L. , T. W. N. Haine, and D. J. Lea, Ruelle's linear response formula, ensemble adjoint schemes and Lévy flights, Nonlinearity 17, 1867-1890 (2004)

[27] Wang, W., Forward and adjoint sensitivity computation of chaotic dynamical systems, J. Comp. Phys. 235, 1-13 (2013)

[28] Ginelli, F., P. Poggi, A. Turchi, H. Chaté, R. Livi, and A. Politi, Characterizing Dynamics with Covariant Lyapunov Vectors, Phys. Rev. Lett. 99, 130601 (2007)

[29] Schubert, S. and V. Lucarini, V., Covariant Lyapunov vectors of a quasi-geostrophic baroclinic model: analysis of instabilities and feedbacks. Q.J.R. Meteorol. Soc., 141, 3040–305 (2015)

[30] Schubert, S. and V. Lucarini, V., Dynamical Analysis of Blocking Events: Spatial and Temporal

Fluctuations of Covariant Lyapunov Vectors, Q.J.R. Meteorol. Soc. doi: 10.1002/qj.2808 (2016)

[31] Lucarini, V., Evidence of dispersion relations for the nonlinear response of the Lorenz 63 system, J. Stat. Phys. 134, 381-400 (2009)

[32] Lucarini, V. and S. Sarno, A statistical mechanical approach for the computation of the climatic response to general forcings, Nonlinear Processes in Geophysics 18, 7-28 (2011)

[33] Lucarini, V., Stochastic Perturbations to Dynamical Systems: A Response Theory Approach. J. Stat. Phys. 146, 774-786 (2012)

[34] Ragone, F., V. Lucarini, and F. Lunkeit, A new framework for climate sensitivity and prediction: a modelling perspective, Climate Dynamics 46, 1459-1471 (2016)

[35] Lucarini, V., F. Lunkeit, and F. Ragone, Predicting Climate Change using Response Theory: Global Averages and Spatial Patterns, J. Stat. Phys. doi: 10.1007/s10955-016-1506-z (2016)

[36] Auerbach, D., P. Cvitanovic, J.-P. Eckmann, G. Gunaratne, I. Procaccia, Exploring chaotic motion through periodic orbits, Phys. Rev. Lett. 58, 2387–2389 (1987)

[37] Cvitanovic, P., Invariant Measurement of Strange Sets in Terms of Cycles, Phys. Rev. Lett. 24, 2729 (1988)

[38] Cvitanovic, P., and B. Eckhardt, Periodic orbit expansions for classical smooth flows, Journal of Physics A 24 L237 (1991)

[39] Cvitanovic, P., R. Artuso, R. Mainieri, G. Tanner and G. Vattay, Chaos: Classical and Quantum, ChaosBook.org (Niels Bohr Institute, Copenhagen 2016)

[40] Gritsun, A., Unstable periodic trajectories of a barotropic model of the atmosphere, Russian Journal of Numerical Analysis and Mathematical Modelling 23, 345–367 (2008)

[41] Gritsun, A., Statistical characteristics, circulation regimes and unstable periodic orbits of a barotropic atmospheric model. Phil. Trans. R. Phil. Soc. A 371, 20120336-20120336 (2012)

[42] Gritsun A., Connection of periodic orbits and variability patterns of circulation for the barotropic model of atmospheric dynamics, Doklady Earth Sciences 438, 636–640 (2011)

[43] Bauer, F., Extended-range weather forecasting. Compendium of Meteorology, T. F. Malone, Ed., Amer. Meteor. Soc., 814–833 (1951)

[44] Rheinhold, B. B., and R. T. Pierrehumbert, Dynamics of weather regimes: Quasi-stationary waves and blocking. Mon. Wea. Rev. 110, 1105–1145 (1982)

[45] Legras, B., and M. Ghil, Persistent anomalies, blocking and variations in atmospheric predictability. J. Atmos. Sci.,42, 433–471(1985)

[46] Benzi R., P. Malguzzi, A. Speranza, and A. Sutera, The statistical properties of general atmospheric circulation: Observational evidence and a minimal theory of bimodality, Q. J. Roy. Met. Soc. 112, 661-674 (1986)

[47] Ghil M. and A. W. Robertson, "Waves" vs. "particles" in the atmosphere's phase space: A pathway to long-range forecasting? PNAS 99, 2493-2500 (2002)

[48] Ghil, M., M. R. Allen, M. D. Dettinger, K. Ide, D. Kondrashov, M. E. Mann, A. W. Robertson, A. Saunders, Y. Tian, F. Varadi, P. Yiou, Advanced spectral methods for climatic time series. Reviews of Geophysics 40, 1003 (2002)

[49] Mukhin, D., A. Gavrilov, A. Feigin, E. Loskutov, and J Kurths, Principal nonlinear dynamical modes of climate variability, Scientific Reports 5, Article number: 15510 (2015)

[50] Selten F.M., Branstator, G., Preferred regime transition routes and evidence for unstable periodic orbit in a baroclinic model, J. Atmos. Sci. 61, 2267-2282 (2004)

[51] Pedlosky, J., Geophysical Fluid Dynamics, Springer, New York 1987

[52] J. Holton, Introduction to Dynamical Meteorology, Academic Press, New York, 2004

[53] D. Ruelle, Chaotic Evolution and Strange Attractors (Cambridge University Press, Cambridge, 1989).

[54] L. Barreira, Y. Pesin, Y., and J. Schmeling, Dimension and product structure of hyperbolic measures. Ann. Math. 149, 755 (1999)

[55] C. L. Wolfe and R. M. Samuelson, An efficient method for recovering Lyapunov vectors from singular vectors, Tellus A 59, 355–366 (2007)

[56] G. Froyland, T. Huls, G.P. Morriss, T.M. Watson. Computing covariant Lyapunov vectors, Oseledets vectors, and dichotomy projectors: A comparative numerical study Physica D 247, 18–39 (2013)

[57] Butterley, O., Liverani, C.: Smooth Anosov flows: correlation spectra and stability. J. Mod. Dyn. 1(2), 301–322 (2007)

[58] H. M. Nussenzveig, Causality and Dispersion Relations, Academic Press, New York, 1972

[59] V. Lucarini. J. Saarinen, K.-E. Peiponen, E. M. Vartiainen, Kramers-Kronig Relations in Optical Materials Research, Springer, New York, 2005

[60] G. J. Chandler and R. R. Kerswell, Invariant recurrent solutions embedded in a turbulent two-dimensional Kolmogorov flow, J. Fluid Mech. 722, 554-595 (2013)

[61] Eckhardt, B. and G. Ott, Periodic orbit analysis of the Lorenz attractor, Z. Physik B - Condensed Matter 93, 259-266 (1994).

[62] Franzke, A. Majda, and E. Vanden–Eijnden, Low-order stochastic mode reduction for a realistic barotropic model climate. J. Atmos. Sci. (2005) 62, 1722–1745.

[63] Kalnay, E., et al., The NCEP/NCAR 40-year reanalysis project. Bull. Amer. Meteor. Soc. 77, 437-471 (1996)

[64] G. Benettin, L. Galgani, A. Giorgilli, J. M. Strelcyn, J. M., Lyapunov Characteristic Exponents for smooth dynamical systems and for hamiltonian systems; A method for computing all of them. Part 2: Numerical application, *Meccanica* 15, 21-30 (1981)

[65] M. Ghil, M. D. Chekroun, and E. Simonnet. Climate dynamics and Fluid mechanics: Natural variability and related uncertainties. Physica D 237, 2111-2126 (2008)

[66] M. D. Chekroun, E. Simonnet, and M. Ghil. Stochastic climate dynamics: Random attractors and time-dependent invariant measures. Physica D 240, 1685-1700 (2011)

[67] A. N. Carvalho, J. Langa, and J. C. Robinson. The pullback attractor. In Attractors for in_nite-dimensional non-autonomous dynamical systems, volume 182 of Applied Mathematical Sciences, 3-22, Springer New York, 2013

[68] P. V. Kuptsov, U. Parlitz, Theory and Computation of Covariant Lyapunov Vectors, Journal of Nonlinear Science 22, 727-762 (2012)

[69] Abramov R, Majda A. J., A new algorithm for low-frequency climate response. J. Atmos. Sci. 66, 286–309 (2009)

[70] Gritsun A, Dymnikov V Barotropic atmosphere response to small external actions. theory and numerical experiments. Izv. Atmos. Ocean. Phys. 35, 511–525 (1999)

[71] von Storch H. and F. Zwiers, Statistical Analysis in Climate Research, Cambridge University Press, Cambridge, 1999

[71] Barnett, T., Intercation of the monsoonand Pacific trade wind system at interannual time scales. Part I: The equatorial zone. Mon. Wea. Rev., 111, 756-773 (1983)

[72] Franzke, C., O'Kane, T. J., Berner, J. B., Williams, P. D., Lucarini, V., Stochastic climate theory and modeling. WIREs Climate Change, 6(1), 63-78 (2015)